\documentclass[fleqn,twoside]{article}
\usepackage{espcrc2}

\usepackage{graphicx}

\newcommand{\be}{\begin{equation}}
\newcommand{\ee}{\end{equation}}

\title{
A Method to Unambiguously Determine the Parity of the $\Theta^+$
Pentaquark}

\author{
A.~W.~Thomas\address[CSSM]{Special Research Centre for the
                       Subatomic Structure of Matter,		\\
                       University of Adelaide,
                       Adelaide SA 5005  Australia}
K.~Hicks\address[ATHENS]{Department of Physics and Astronomy, \\
Ohio University, Athens, Ohio 45701 USA}$^{\rm ,c}$,
A.~Hosaka\address[RCNP]{Research Center for Nuclear Physics (RCNP), \\ 
Osaka University, Ibaraki 567-0047, Japan},
}

\begin{document}

\begin{abstract}
With the recent discovery of the $\Theta(1540)$ pentaquark,
the question of its parity is paramount since this will
constrain the correct description of its internal structure.
We show that the measurement of the spin singlet and triplet 
cross sections for the reaction $\vec{p}\vec{p} \to \Sigma^+ \Theta^+$ 
will unambiguously determine the parity of the $\Theta^+$. 
\end{abstract}


\maketitle
The recent discovery of the $\Theta^+$ 
baryon~\cite{Nakano:2003qx,Barmin:2003vv,Stepanyan:2003qr} 
has been associated with a tremendous amount of theoretical 
activity~\cite{jlab}. For the present there is no consensus 
within the theoretical community as to the parity of this exotic state, with 
roughly half of the calculations/models on either side of the mirror.
For example, the original prediction of the $\Theta^+$ in the chiral 
soliton model \cite{Diakonov:1997} predicted that the $\Theta^+$ (called 
the $Z^+$ in that paper) is part of an anti-decuplet, with all members 
having spin and parity $J^\pi = \frac{1}{2}^+$.  On the other hand, 
lattice gauge calculations \cite{lattice:2003} suggest that the 
$\Theta^+$ has $J^\pi = \frac{1}{2}^-$.
Since the parity reflects internal dynamics of the $\Theta^+$, 
it is absolutely crucial to further theoretical progress that the parity be 
determined experimentally, as soon as possible and without ambiguity. 

A number of proposals have been 
made so far~\cite{theory}, but all of them involve considerable experimental 
challenge as well as an understanding of the reaction mechanism. 
There is good reason why this is so difficult.  Suppose that a sample of 
100\% polarized $\Theta^+$ particles could be prepared at rest in the 
laboratory.  Even under this ideal condition, the decay angular distribution 
of this strongly-decaying particle gives information only on the 
spin, and not the parity, unless the polarization of the final-state 
nucleon is measured (this is a simple consequence of symmetry of the 
magnetic substates of the system)~\footnote{
In fact, the angular distribution of the kaon from a polarised $\Theta^+ 
(J_z =J)$ is proportional to $(\sin \theta)^{2J-1}$ regardless the parity}.
Even if the difficult experimental 
task of measuring the final nucleon polarization is accomplished for the 
small $\Theta^+$ cross section, the polarization of the $\Theta^+$ 
depends on the density matrix elements of the reaction mechanism. It would be 
much more desirable to have a method that is independent of the details of 
the production mechanism.

A classic, well-known example of parity determination is the case of the 
pion, which involved the decay of pionic deuterium~\cite{panofsky}.   
There the Fermi statistics of the two nucleons and the threshold
kinematics played an essential role.  
Here we consider an analogous reaction, namely $\Theta^+$ production 
from two protons in the threshold region.  We will briefly 
explain how this process can provide an unambiguous determination of the 
parity of the $\Theta^+$. For the purposes of our analysis we assume that the
$\Theta^+$ has spin-1/2, but it is trivial to show that the argument 
generalises to spin-3/2. We consider the process
\begin{equation}
\vec p + \vec p \rightarrow \Sigma^+ + \Theta^+ \, ,
\label{eq:one}
\end{equation}
at and just above threshold. If the final centre of mass momentum is $k$, 
a final state with $L=1$ is suppressed in production rate with respect 
to $L=0$ by a factor 
$(kR)^2$, with $R$ a characteristic distance of order one fermi. If one is 
within a few MeV of threshold the suppression is one to two orders of 
magnitude. Thus one can be sure that the final state has $L=0$. (The energy 
dependence of the production cross section in the region just above threshold 
will, in any case, provide an unambiguous check of this assumption.) The total 
spin of the $\Sigma^+$ and the $\Theta^+$ is $S$ = 0 or 1. Thus the total 
angular momentum and parity of the final state must be $0^+$ or $1^+$ if the 
parity of the $\Theta^+$ is positive and $0^-$ or $1^-$ if the 
parity of the $\Theta^+$ is negative. Since the total angular momentum and 
parity are conserved in strong interactions these values will also be the 
total angular momentum and parity of the initial state.    

We note that the isospin of the initial $pp$ state is $I=1$ and the Pauli 
exclusion principle then implies that if the initial $pp$ orbital angular 
momentum is even the initial spin must be $S=0$, 
while if it is odd the initial 
spin must be $S=1$. This leads to the following possibilities:
\begin{itemize}
\item {\bf Parity of $\Theta^+$ positive:}
In this case, the spins of the protons should be
anti-alligned.
Then the initial $pp$ state must be $^1S_0$.
\item {\bf Parity of $\Theta^+$ negative:}
In this case, the spins of the protons should be
alligned.
Then the initial $pp$ state must be $^3P_0$ or $^3P_1$ 
(for $J=0$ and 1, respectively).
\end{itemize}
No other possibilities exist.

Clearly a measurement with polarised proton beam and target which observes 
the reaction~(\ref{eq:one}) near threshold enables 
one to determine whether the 
production occurs for the spin singlet or triplet state of the initial 
protons. If it is the former the $\Theta^+$ {\bf must} have positive parity, 
while if the latter it {\bf must} have negative parity. We stress that this 
conclusion relies only on the conservation of total angular momentum, parity 
and isospin in the strong interaction and is totally independent of any 
particular reaction mechanism. 

Assuming that the width of the $\Theta^+$ is of the order an MeV or more, 
one would expect the total cross section for the reaction~(\ref{eq:one}) 
to be in the range 0.1 to 1.0 $\mu$b at beam energies a few tens of MeV 
above the reaction threshold \cite{COSY}. Near threshold, the cross section 
will be lower, perhaps by one or two orders of magnitude.  In addition, the 
requirement of a polarized beam and a polarized target could reduce the 
luminosity of the measurement, making it still more difficult.  However, 
near-threshold measurements are now routinely done with polarized beam 
and target \cite{IUCF} that were considered extremely difficult in the 
previous years.  
Considering the overwhelming theoretical importance of determining the
parity of the $\Theta^+$, experimentalists have a strong motivation to
use their ingenuity and overcome the difficulties of such a measurement.
We expect that it is within the 
the capabilities of facilities such as COSY at J\"ulich, and it is important 
that this measurement be carried out as soon as possible.

\section*{Acknowledgments}
We thank Takashi Nakano for discussions and comments. 
AWT and KH would like to acknowledge the hospitality of the 
Research Centre for Nuclear Physics (RCNP) at Osaka University.
This work was supported by the Australian Research Council and 
the U.~S.~National Science Foundation.

\end{document}